\documentclass[aps,prd,10pt,twocolumn,nofootinbib,superscriptaddress]{revtex4}
\usepackage{mathrsfs}  
\usepackage{adjustbox}
\usepackage{epsfig, cancel}
\usepackage{latexsym}
\usepackage{natbib, comment}
\usepackage{url}
\usepackage{dcolumn}
\usepackage{color}
\usepackage{amsfonts,amssymb,amsmath, txfonts}
\usepackage{graphicx,epsfig}
\usepackage{psfrag}
\usepackage{subfigure}
\usepackage{hyperref}
\hypersetup{colorlinks=true}
\usepackage{mathtools}
\usepackage{enumitem}
\usepackage{float}
\usepackage[dvipsnames]{xcolor}
\usepackage{xcolor}
\hypersetup{ linktoc=all,
    colorlinks, linkcolor={brightpink},
    citecolor={blue}, urlcolor={blue}
}
\definecolor{rosy}{RGB}{230,235,252}
\definecolor{myframetitle}{RGB}{90,89,170}
\definecolor{myblocktitle}{RGB}{140,185,249}
\definecolor{mytitle}{RGB}{10,80,26}

\definecolor{darkgreen}{RGB}{27,130,45}
\definecolor{darkblue}{rgb}{0,0,0.3}
\definecolor{darkred}{rgb}{0.7,0,0}

\definecolor{light gray}{RGB}{220,220,220}
\definecolor{dark purple}{RGB}{108,0,217}
\definecolor{pink}{RGB}{190,20,100}
\definecolor{orang}{RGB}{193,63,0}
\definecolor{green}{RGB}{11,98,17}
\definecolor{darkpink}{RGB}{153,0,76}
\definecolor{bluegreen}{RGB}{0,102,102}
\definecolor{greenlagan}{RGB}{0,102,0}
\definecolor{redgreen}{RGB}{102,102,0}
\definecolor{Redgreen}{RGB}{153,76,0}
\definecolor{vividviolet}{rgb}{0.62, 0.0, 1.0}
\definecolor{amaranth}{rgb}{0.9, 0.17, 0.31}
\definecolor{palatinateblue}{rgb}{0.15, 0.23, 0.89}
\definecolor{brightpink}{rgb}{1.0, 0.0, 0.5}
\definecolor{cornflowerblue}{rgb}{0.39, 0.58, 0.93}
\definecolor{deepcarminepink}{rgb}{0.94, 0.19, 0.22}
\definecolor{radicalred}{rgb}{1.0, 0.21, 0.37}

\def\H0{{\text{H}\hspace*{-2.05mm}\text{H} 0\hspace*{-1.35mm}0\ }}

\def\be{\begin{equation}}
\def\ee{\end{equation}}
\def\beq{\begin{equation}}
\def\eeq{\end{equation}}
\def\bea{\begin{eqnarray}}
\def\eea{\end{eqnarray}}

\newcommand{\nn}{\nonumber \\}

\begin{document}

\title{Elucidating cosmological model dependence with $H_0$}

 \author{Eoin \'O Colg\'ain}\email{ocolgain@gmail.com}
 \affiliation{Center for Quantum Spacetime, Sogang University, Seoul 121-742, Korea}
 \affiliation{Department of Physics, Sogang University, Seoul 121-742, Korea} 
 \author{M. M. Sheikh-Jabbari}\email{shahin.s.jabbari@gmail.com}
\affiliation{School of Physics, Institute for Research in Fundamental Sciences (IPM), P.O.Box 19395-5531, Tehran, Iran}

\begin{abstract}
We observe that the errors on the Hubble constant $H_0$, a universal parameter in any FLRW cosmology, can be larger in specific cosmological models than Gaussian Processes (GP) data reconstruction. We comment on the prior mean function and trace the smaller GP errors to stronger correlations, which we show precludes all well studied dynamical dark energy models.    
Our analysis suggests that ``cosmological model independence", especially in the statement of Hubble tension, has become a misnomer.
\end{abstract}

\maketitle

\section{Introduction}
Cosmology rests upon assumptions. When one works with assumptions, timely contradictions are inevitable and these seed progress. Over two decades ago, the concordance flat $\Lambda$CDM model emerged from a set of contradictions. Today, cosmological tensions \cite{Verde:2019ivm} point to problems with the \textit{assumption} that the Universe is flat $\Lambda$CDM. Moreover, some \textit{assumptions} underlying supernovae are in a state of flux \cite{Kang:2019azh, Rose:2020shp, Brout:2020msh}, and even the \textit{assumption} that the Universe is isotropic \& homogeneous is being called into question \cite{Secrest:2020has, Migkas:2021zdo}. This perpetual cycle of assumptions and contradictions is integral to cosmology. Recently, Gaussian Processes (GP) has become a staple of data-driven cosmology \cite{Holsclaw:2010nb, Holsclaw:2010sk, Shafieloo:2012ht, Seikel:2012uu, GP-H0-value, GP-DE-EoS, GP-Omegak}.  In this letter, {using the Hubble constant $H_0$, we chip away} at the widespread assumption that GP data reconstruction  is cosmological model independent. 

Cosmology strives to make robust statements across a host of cosmological models and this drives the ``model independence" narrative. Working within parametric models, it is well established that Taylor expansion, or cosmography \cite{Visser:2004bf, Visser-TE}, offers one a glimpse of model independence, but the Cauchy-Hadamard theorem \cite{CH} (see \cite{Cattoen:2007sk}) confines one to low redshifts, $ z \lesssim 1$. Nevertheless, even within these restrictions, the Hubble constant $H_0$ can be determined in a \textit{bona fide} model independent manner \cite{Riess:2016jrr, Dhawan:2020xmp}. In contrast, GP reconstruction is a non-parametric technique that in principle allows one to extend ``model independence" to higher redshifts. In practice, one reconstructs data through an assumption on the covariance matrix or kernel and its ``hyperparameters". 

The commonly held belief that GP is model independent {may be misleading} for largely two reasons. First, cosmological inferences of $H_0$ from GP \cite{Busti:2014dua, Yu:2017iju, Gomez-Valent:2018hwc, Haridasu:2018gqm, Bonilla:2020wbn, Renzi:2020fnx} at the percent level can be discrepant with local $H_0$ determinations \cite{Riess:2019cxk} (see also \cite{DiValentino:2020vnx}).\footnote{These percent level cosmological model independent $H_0$ determinations largely leave one questioning the systematics.} If true, it is an immediate corollary that Hubble tension has no cosmological resolution, at least within the Friedmann-Lema\^itre-Robertson-Walker (FLRW) framework. Admittedly, this may be true, so there is no contradiction. Nevertheless, more seriously, Table \ref{nonparam} shows the average errors for $300+$ flat $\Lambda$CDM mock realisations with forecasted DESI data \cite{Aghamousa:2016zmz}, where GP based on commonly used kernels in the Mat\'ern covariance matrix class with positive parameter $\nu$ (e.g. see \cite{Seikel:2013fda}) is compared against the ubiquitous Chevallier-Polarski-Linder (CPL) model \cite{Chevallier:2000qy, Linder:2002et} for dynamical dark energy (DDE). As we argue later, similar results should hold for all parametric DDE models. The obvious question is how does a putative ``model independent" technique outperform a specific model on errors? Recall that typically, where there's smoke, there's fire. 

\begin{table}[htb]
\centering
\begin{tabular}{c|ccccc}
\rule{0pt}{3ex} Model & $\nu=\infty$ & $\nu=9/2$ & $\nu=7/2$ & $\nu=5/2$  & CPL \\
\hline
\rule{0pt}{3ex} $H_0$  &  $68.59^{+2.21}_{-2.20}$ & $68.49^{+2.34}_{-2.34}$  & $68.42^{+2.45}_{-2.45}$ &  $68.44^{+2.73}_{-2.73}$ &  $69.54^{+3.30}_{-3.37}$\\
\end{tabular}
\caption{Average values of $H_0$ for different models based on between 300+ mock realisations of forecasted DESI data. }
\label{nonparam}
\end{table}

In essence, the overt problem with non-parametric techniques, such as GP, is that the implications for parametric models are covert. Obviously, in its simplest setting with observational Hubble data (OHD), one assumes a kernel, either extremises or marginalises over hyperparameters in a likelihood and outputs a mean $H(z_i)$ at redshifts $z_i$, as well as an associated covariance matrix.  In contrast, when one fits a parametric model, one recovers the best-fit parameters and their covariance, typically from a Markov Chain Monte Carlo (MCMC) chain. In principle, one can infer $H(z_i)$ from the MCMC chain and this facilitates a direct comparison. Alternatively, one can strip away the errors in $H(z_i)$ in both cases and directly compare the correlation matrix. 

In this note, we focus on $H_0$, which, as remarked in \cite{Krishnan:2020vaf}, is an integration constant in the Friedmann equations, so it is universal to all FLRW cosmologies.  Concretely, we show that while the correlations in simpler models such as flat $\Lambda$CDM and $w$CDM are typically stronger than the GP output, in turn correlations from GP are generically stronger than DDE models. As a result, GP represents a restriction on the parameter space of DDE models. This explains the smaller errors in Table \ref{nonparam} and highlights the problem with the assumption that GP is model independent. 

\section{Taylor expansion}
\label{sec:TE}
We warm up by putting to bed commonly propagated misconceptions regarding cosmography \cite{Visser:2004bf}, {which may or may not echo previous studies in this direction \cite{Expansions}.} Let us begin with two relevant math theorems (see for example \cite{CH, Spivak}). 

\textbf{Taylor's Theorem:} \textit{Let $n \geq 1$ be an integer and the let the function $f: \mathbb{R} \rightarrow \mathbb{R}$ be $n$ time differentiable the the point $z_0 \in \mathbb{R}$. Then there exists a function $f_{n}: \mathbb{R} \rightarrow \mathbb{R}$ such that 
\bea
\label{taylor_exp}
f(z) &=& f(z_0) + f'(z_0) (z-z_0) + \frac{f''(z_0)}{2!}  (z - z_0)^2 + \dots \nn 
&+& \frac{f^{(n)}}{n!}  (z_0) (z-z_0)^n + f_{n} (z) (z-z_0)^n, 
\eea
and $\lim_{z \rightarrow z_{0}} f_n(z) = 0$. } 

\textbf{Cauchy-Hadamard Theorem:} \textit{Consider the formal power series in $z \in \mathbb{C}$ of the form $f(z) = \sum_{n=0}^{\infty} c_n (z - z_0)^{n}$, where $z_0, c_{n} \in \mathbb{C}$. Then the radius of convergence $R$ of $f$ at the point $z_0$ is given by }
\be
\label{CH}  \frac{1}{R} = \limsup_{n \rightarrow \infty} ( |c_n|^{1/n} ). 
\ee

Observe that Taylor's theorem simply guarantees that provided the Hubble parameter $H(z)$ is differentiable, which is usually the case, the remainder function $f_n(z)$ exists and approaches zero as $z$ approaches $z_0$. While one could perform this expansion at any redshift, it is natural to consider expansions around $z = 0$, and this is the basis for cosmographic (Taylor) expansions \cite{Visser:2004bf, Visser-TE}. Note, working in the vicinity of $z=0$ is also sufficient for determining $H_0$. 

Before proceeding, a comment on ``model independence" of Taylor expansions is in order. Recalling the earlier discussions, one can be confident that in the \textit{immediate} vicinity of $z_0$ \textit{all} models are covered, so Taylor expansion is model independent in a real sense. This is essentially the regime that Riess et al. \cite{Riess:2016jrr} operate in to make local determinations of $H_0$ (see also \cite{Dhawan:2020xmp}). The farther one goes from $z=z_0$, the fewer models that are accurately described by the Taylor expansion and it only covers a \textit{class of models}. 

As observed in \cite{Visser-TE}, recalling the definition of the Hubble parameter in terms of the scale factor, $H \equiv \dot{a}/a$, and the usual expression for $a$ in terms of redshift $z$, $a = 1/(1+z)$, it is clear that the scale factor becomes singular at $z = -1$. We can extend $z$ into the complex plane, where in the Cauchy-Hadamard language (\ref{CH}), this singularity corresponds to at least one of the $c_n$ becoming large at $z \approx -1$. This in turn ensures $R \rightarrow 0$ in its vicinity. For this reason, as stated in \cite{Visser-TE}, the radius of convergence of any FLRW cosmology is \textit{at most} $|z| =1$. It should be clear that {Taylor's theorem does not apply to expansions in $(1+z)$ about $z=0$, e. g.  \cite{Sahni:2002fz, Sahni:2006pa}, only expansions about $z=-1$, and there the radius of convergence is strictly zero. Together, these theorems make expansions in $(1+z)$, or $\log_{10} (1+z)$ \cite{Lusso:2019akb} completely random in the sense that adding higher order terms does not improve convergence (see \cite{Yang:2019vgk, Banerjee:2020bjq}). Further comments can be found in appendix \ref{moreTE}.}

\section{Gaussian Processes}\label{sec:GP}
GP is a method to smooth a given (sparse) dataset. In essence, given $n$ observational real data points $\mathbf{y} = (y_1, \dots y_n)$ at redshifts $\mathbf{z} = (z_1, \dots z_{n})$ with a covariance matrix $C$, one wishes to reconstruct a function $\mathbf{f}^* = (f(z_1^*), \dots f(z_N^*))$ underlying the data at $N$ new points $\mathbf{z}^{*} = (z_1^*, \dots, z_{N}^*)$, where typically 
$N \geq n$. Obviously, attempting to reconstruct data far outside the range of the original data will lead to questionable results.\footnote{That being said, we will be taking a slight liberty with the range of the original data in extrapolating from $z \sim 0.07$ down to $z \sim 0$ to extract $H_0$. This is in line with the analysis of \cite{Gomez-Valent:2018hwc}.}

In implementing GP, one has to make an assumption on how the reconstructed data points are correlated, and to do so, one introduces a new covariance matrix $K(\mathbf{z}^*, \mathbf{z}^*)$, typically called a ``kernel". The kernel $K$ is a function of some hyperparameters, in cosmological applications commonly taken to be two  ($\sigma_f, \ell_f)$. The most commonly used kernel, from which the method derives its name, is Gaussian,  
\be
\label{gaussian}
K (z, \tilde{z}) = \sigma_f^2 \exp \left(- \frac{(z-\tilde{z})^2}{2 \ell_f^2} \right). 
\ee
The other  kernels that are commonplace in cosmological settings are the Mat\'ern covariance functions, e.g. see \cite{Seikel:2013fda},
\be
\label{matern}
K_{\nu}(z, \tilde{z}) = \sigma_f^2 \frac{2^{1-\nu}}{\Gamma(\nu)} \left( \frac{\sqrt{2 \nu (z- \tilde{z})^2 }}{\ell_f}\right)^{\nu} \tilde{K}_{\nu} \left( \frac{\sqrt{2 \nu (z- \tilde{z})^2 }}{\ell_f }\right), 
\ee
where $\Gamma$ is the gamma function and $\tilde{K}_{\nu}$ is a modified Bessel function. Here $\nu$ is a positive parameter and in the  $\nu \rightarrow \infty$ limit  one recovers the Gaussian kernel (\ref{gaussian}). It should be noted that the Mat\'ern kernels are only mean square $n$-differentiable provided $\nu > n$.  This differentiability property is important when one is interested in the derivatives of $H(z)$, but as we work here with OHD, this is less of a concern. In addition to the Gaussian, following \cite{Seikel:2013fda}, we will largely focus on $\nu = p +\frac{1}{2}$, where $p=0, 1, 2, 3, 4$ (see appendix \ref{App:Matern}). 

Since we are only interested in $H(z)$, the mean $\overline{f^*}$ and the covariance cov($f^*)$ from the GP reconstruction can be easily constructed through a few lines of linear algebra \cite{Seikel:2012uu}: 
\be\label{f_cov}\begin{split}
\overline{f^*} =& \mathbf{\mu}(\mathbf{z}^*) + K(\mathbf{z}^*, \mathbf{z}) [ K(\mathbf{z}, \mathbf{z}) + C]^{-1} \left( \mathbf{y} - \mathbf{\mu}(\mathbf{z}) \right), \\
\textrm{cov}{(f^*)} =& K(\mathbf{z}^*, \mathbf{z}^*) - K(\mathbf{z}^*, \mathbf{z}) [ K(\mathbf{z}, \mathbf{z}) + C]^{-1} K(\mathbf{z}, \mathbf{z}^*), 
\end{split}\ee
where $\mathbf{\mu}(\mathbf{z})$ is a prior mean function that one commonly sets to zero, $\mathbf{\mu}(\mathbf{z}) = 0$ \cite{Seikel:2012uu} (see also \cite{Busti:2014dua, Yu:2017iju, Gomez-Valent:2018hwc, Haridasu:2018gqm, Bonilla:2020wbn, Renzi:2020fnx}). 

The only problem now is to identify the hyperparameters and this is done through the following log normal likelihood:
\be\label{likelihood}\begin{split}
\ln \mathcal{L} &=- \frac{1}{2} (\mathbf{y} - \mu(\mathbf{z}))^{T} [ K(\mathbf{z}, \mathbf{z}) + C]^{-1} (\mathbf{y} - \mu(\mathbf{z})) \\
&- \frac{1}{2} \ln | K(\mathbf{z}, \mathbf{z}) +C| - \frac{n}{2} \ln 2 \pi. 
\end{split}\ee
In a strict Bayesian sense, one should marginalise over the hyperparameters through an MCMC routine, e. g. \cite{ForemanMackey:2012ig}. However, this is computationally more expensive, so in practice it is common to simply optimise (\ref{likelihood}), by setting to zero the gradient of $\ln \mathcal{L}$,
\be\label{grad}
\begin{split}
\nabla (\ln \mathcal{L} ) =& \frac{1}{2} (\mathbf{y} - \mu) ^{T} (K + C)^{-1} \nabla K ( K + C)^{-1} ( \mathbf{y} - \mu(\mathbf{z}) ) \\
 -& \frac{1}{2} \textrm{tr} [ (K+C)^{-1} \nabla K]. 
\end{split}\ee

\subsection{Mean function}
A glance at the literature reveals that the GP community breaks up into two schools. Here, we follow Seikel et al. \cite{Seikel:2012uu}, where the zero mean function, $\mu(z) = 0$, is imposed. This choice is closer to our interests, as it represents the methodology that has led to curiously small errors on $H_0$ \cite{Busti:2014dua, Yu:2017iju, Gomez-Valent:2018hwc, Haridasu:2018gqm, Bonilla:2020wbn, Renzi:2020fnx}. When $\mu(z) = 0$, irrespective of whether one optimises (\ref{likelihood}) or marginalises over the hyperparameters, there is very little difference to the results \cite{Seikel:2013fda, Gomez-Valent:2018hwc}. As is clear from (\ref{likelihood}) or (\ref{grad}), since $\mathbf{y}^2 \gg \sigma_{\mathbf{y}}^2 \sim C$, the hyperparameter $\sigma_f$ is a large number. As a result, $K \gg C$, which is clear from the values in Table \ref{H0kernel}. With this difference in scales, one can approximate 
\be
(K + C)^{-1} \approx K^{-1} - K^{-1} C K^{-1} + \dots, 
\ee
and the mean and covariance matrix become to leading order: 
\be\label{approx1}\begin{split}
\overline{f^*} &=  {\cal D}(\mathbf{z}^*, \mathbf{z}) \ \mathbf{y}  + \dots, \\
\textrm{cov}{(f^*)} &= {\cal D}(\mathbf{z}^*, \mathbf{z})\ C\  {\cal D}( \mathbf{z}, \mathbf{z}^*) 
+ \dots, 
\end{split}\ee
where ${\cal D}(\mathbf{z}^*, \mathbf{z}):=K(\mathbf{z}^*, \mathbf{z}) K(\mathbf{z}, \mathbf{z})^{-1}$ is the ``dressing matrix''
which essentially dresses the original data $\mathbf{y}$ and covariance matrix $C$. It is a matrix of $O(1)$ numbers. It should be clear from the leading order expressions that GP implemented with zero mean $\mu(\mathbf{z})$ is a mapping from data $\mathbf{y}$ into the mean $\overline{f^*}$, and a mapping from the covariance matrix $C$ into the reconstructed covariance matrix  $\textrm{cov}{(f^*)}$. 

The other school, primarily Shafieloo et al. \cite{Shafieloo:2012ht}, maintains that the prior on the mean is important. As is clear from (\ref{likelihood}) or (\ref{grad}), a reasonably competitive guess for the mean should lead to a small $\mathbf{y} - \mu(\mathbf{z})$, which makes $\sigma_f$ a small number, $\sigma_f \sim K \sim O(1)$. For this reason, one just recovers the input mean if one optimises the likelihood (\ref{likelihood}) and one \textit{must} marginalise over the hyperparameters. This marks a key distinction between the two approaches. The other important difference is that one expects marginalisation to lead to a distribution of $\sigma_f$ peaked in the vicinity of $\sigma_f \approx 0$.  Therefore,  one is working in the opposite regime to (\ref{approx1}) where now $C \gg K$. In this case, the mean and the covariance  to leading {and sub-leading} order are,
\be\label{approx2}\begin{split}
\overline{f^*} =&  \mu(\mathbf{z}^*) + K(\mathbf{z}^*, \mathbf{z}) C^{-1} ( \mathbf{y} - \mu(\mathbf{z}) )   + \dots, \\
\textrm{cov}{(f^*)} =& K(\mathbf{z}^*, \mathbf{z}^*)  - K(\mathbf{z}^*, \mathbf{z})  C^{-1} K(\mathbf{z}, \mathbf{z}^*) + \dots.
\end{split}
\ee 
{It should be clear that these expressions are trivial at leading order: one gets out what one puts in. So, the sub-leading terms have to matter and this is where marginalisation helps. Nevertheless, the better the guess on the mean, which is typically inferred from specific models, the smaller the sub-leading terms, and hence the less relevant the GP becomes.  Clearly, choosing a prior is a balancing act that represents additional modeling in this ``model independent" approach and for this reason, we set $\mu(z)=0$.}

\subsection{Data} 
\begin{figure}[htb]
\centering
\includegraphics[angle=0,width=80mm]{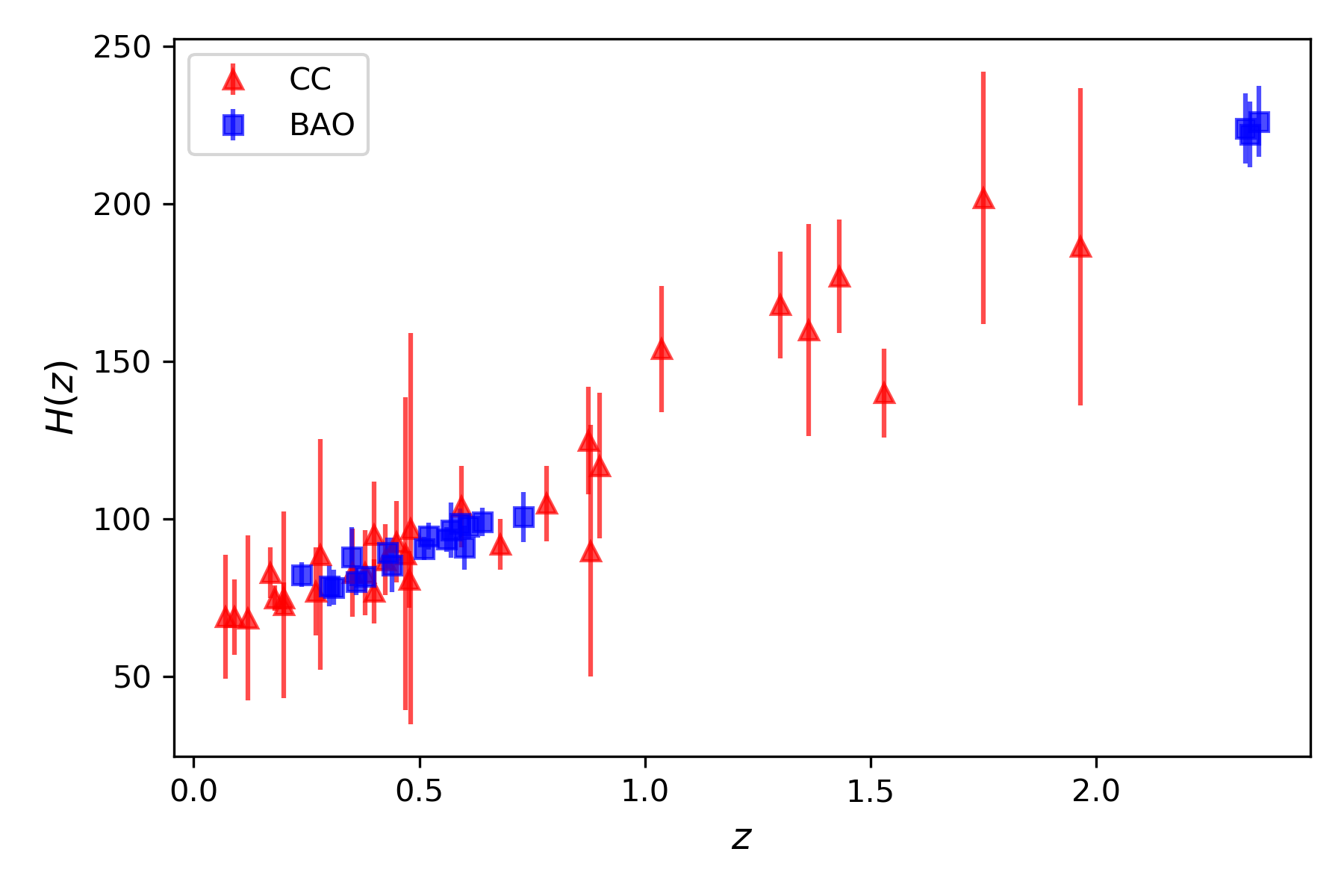} \\
\caption{The real CC and BAO data serving as the basis for mock realisations.}
\label{data}
\end{figure}

We use OHD, which serves as the basis for mock realisations. More precisely, we make use of cosmic chronometer (CC) \cite{Jimenez:2001gg, CC} and homogenised BAO \cite{Eisenstein:2005su, Magana:2017nfs, BAO-data} data. It should be stressed that the CC data largely comprises statistical errors only, and the systematic errors on $H(z)$ are a work in progress \cite{Moresco:2020fbm}. That being said, this OHD will only serve as the basis for mock realisations of the flat $\Lambda$CDM cosmological model with the canonical parameters $(H_0, \Omega_{m0}) = (70, 0.3)$. Furthermore, we are not interested in the absolute value of $H_0$, but the errors and only their relative values. We present  the OHD in FIG. \ref{data}. From the real data, we extract the redshifts $z_i$ and the errors in the Hubble parameter $\sigma_{H(z_i)}$. To perform the mocks,  at each $z_i$ we choose a new $H(z_i)$ value from a normal distribution about the flat $\Lambda$CDM value with standard deviation $\sigma_{H(z_i)}$. 

\section{Analysis}\label{sec:GPA}
In this section we focus on $H_0$ extracted from the GP reconstruction. This is arguably the simplest cosmological parameter that one can reconstruct from the data since it just involves an extrapolation beyond the last data point ($z=0.07$ in our study) to $z = 0$. 
{Before beginning, it is instructive to remove the BAO data and run the GP analysis for the CC data with a Gaussian kernel, just to validate our GP code. We find $H_0  = 67.55^{+4.75}_{-4.68}$ km/s/Mpc, which reproduces the result of Yu, Ratra \& Wang \cite{Yu:2017iju}, $H_0 = 67.42 \pm 4.75$ km/s/Mpc, so there is no indication that our GP code is doing anything unusual. In particular, the errors are the same size. It should be stressed again that GP is simple linear algebra (\ref{f_cov}). 

Here, we begin exploring the GP  whereby the likelihood  (\ref{likelihood}) is minimised. {This represents a simplification, but it has been confirmed in \cite{Seikel:2013fda, Gomez-Valent:2018hwc} that this make little difference.} In Table \ref{H0kernel} we show how the inferred Hubble constant $H_0$ depends on the kernel for the full redshift range of the data $0 \lesssim z \lesssim 2.5$. It should be stressed that we are using OHD, namely CC and BAO data, but since we average over a large number of mocks, we are reporting general trends. We find that errors on $H_0$ decrease with increasing $\nu$, and our analysis shows that the smallest $H_0$ error is achieved for the Gaussian kernel. Our findings are in line with Table 1 of \cite{Busti:2014dua}, where we have included the $\nu = 1/2$ and $\nu = 3/2$ entries just to fill out the picture. Note that results are kernel dependent. 

\begin{table}[htb]
\centering
\begin{tabular}{c|ccc}
\rule{0pt}{3ex} $K_{\nu}$ & $H_0$ (km/s/Mpc) & $\sigma_f$ & $\ell_f$ \\
\hline
\hline 
\rule{0pt}{3ex} $\nu = 1/2$  & $73.92^{+13.23}_{-13.22}$ & $176.52^{+180.72}_{-172.24}$ &  $43.45^{+4.65}_{-5.52}$ \\
\hline 
\rule{0pt}{3ex} $\nu = 3/2$  & $68.61^{+5.11}_{-5.11}$ & $284.23^{+36.08}_{-38.17}$ &  $8.77^{+1.65}_{-1.83}$ \\
\hline 
\rule{0pt}{3ex} $\nu = 5/2$  & $68.81^{+3.91}_{-3.90}$ & $252.51^{+39.03}_{-40.47}$ &  $5.19^{+0.91}_{-0.92}$ \\
\hline 
\rule{0pt}{3ex} $\nu  = 7/2$  & $69.35^{+3.64}_{-3.64}$ & $241.94^{+39.06}_{-38.64}$ &  $4.30^{+0.76}_{-0.80}$ \\
\hline
\rule{0pt}{3ex} $\nu = 9/2$  & $69.25^{+3.57}_{-3.57}$ & $237.34^{+38.00}_{-40.06}$ &  $3.91^{+0.66}_{-0.75}$ \\
\hline
\rule{0pt}{3ex} Gaussian ($\nu = \infty$) & $69.56^{+3.42}_{-3.42}$ & $230.20^{+41.48}_{-42.08}$ &  $3.03^{+0.60}_{-0.57}$ \\
\end{tabular}
\caption{Average values of $H_0$ and hyperparameters ($\sigma_f, \ell_f$) for different kernels and 500 mock realisations of the data.}
\label{H0kernel}
\end{table}

Observe that the central $H_0$ values are all biased lower than the mock value $H_0 = 70$ km/s/Mpc. Independently, we have performed some fits with Taylor expansions and one observes the same phenomenon, which suggests that this biasing is down to the data. As is evident from FIG. \ref{data}, the error bars increase at low redshifts,  the slope of $H(z)$ is fixed by BAO,  making it less likely that the visibly poorer quality CC data can affect the central value.

It is instructive to fit the same data to concrete models in order to compare the errors in $H_0$. The result is reported in Table \ref{param}. Evidently, the errors on $H_0$ for both $\Lambda$CDM and $w$CDM are within the GP errors, but for CPL we find that the error on $H_0$ is larger. In order to ascertain if this is a fluke, we replace our OHD  with the DESI  $H(z)$ determination forecasts in the extended redshift range $0.05 \leq z \leq 3.55$ \cite{Aghamousa:2016zmz},\footnote{See \cite{Bengaly:2019ibu} for how related forecasts will constrain GP constraints on deceleration.}  where we assume the optimistic outcome that the five-year survey covers 14,000 deg$^2$. Repeating the exercise, we can see from Table \ref{nonparam} that even with the forecasted data, GP outperforms the CPL model by leading to smaller errors on $H_0$. We conclude that this is not an artifact of the dataset and that GP produces smaller errors on $H_0$ than CPL. 

\begin{table}[htb]
\centering
\begin{tabular}{c|ccc}
\rule{0pt}{3ex} Model & $\Lambda$CDM & $w$CDM & CPL \\
\hline
\rule{0pt}{3ex} $H_0$ (km/s/Mpc) &  $69.91^{+1.17}_{-1.20}$ & $69.79^{+3.05}_{-2.86}$ & $65.64^{+5.11}_{-4.95}$ \\
\end{tabular}
\caption{Average values of $H_0$ for different models based on $\sim 300$ mock realisations of the data in FIG. \ref{data}. }
\label{param}
\end{table}

\subsection{Correlations} 
Recall that the output from GP is a mean and a covariance matrix and that any covariance matrix $C_{ij}$ is simply a dressing of the correlation matrix $D_{ij}$ through the errors $\sigma_i$, $C_{ij} = \sigma_i  \sigma_j D_{ij} $ {(no summation)}. It is an easy task to take the output covariance matrix from GP and identify the underlying correlation matrix. Concretely, we have 
\be
\textrm{cov}{(f^*)} (z_i, z_j) = \sigma_{H(z_i)} \sigma_{H(z_j)} D ( z_i, z_j). 
\ee

\begin{figure}[htb]
\centering
\includegraphics[angle=0,width=80mm]{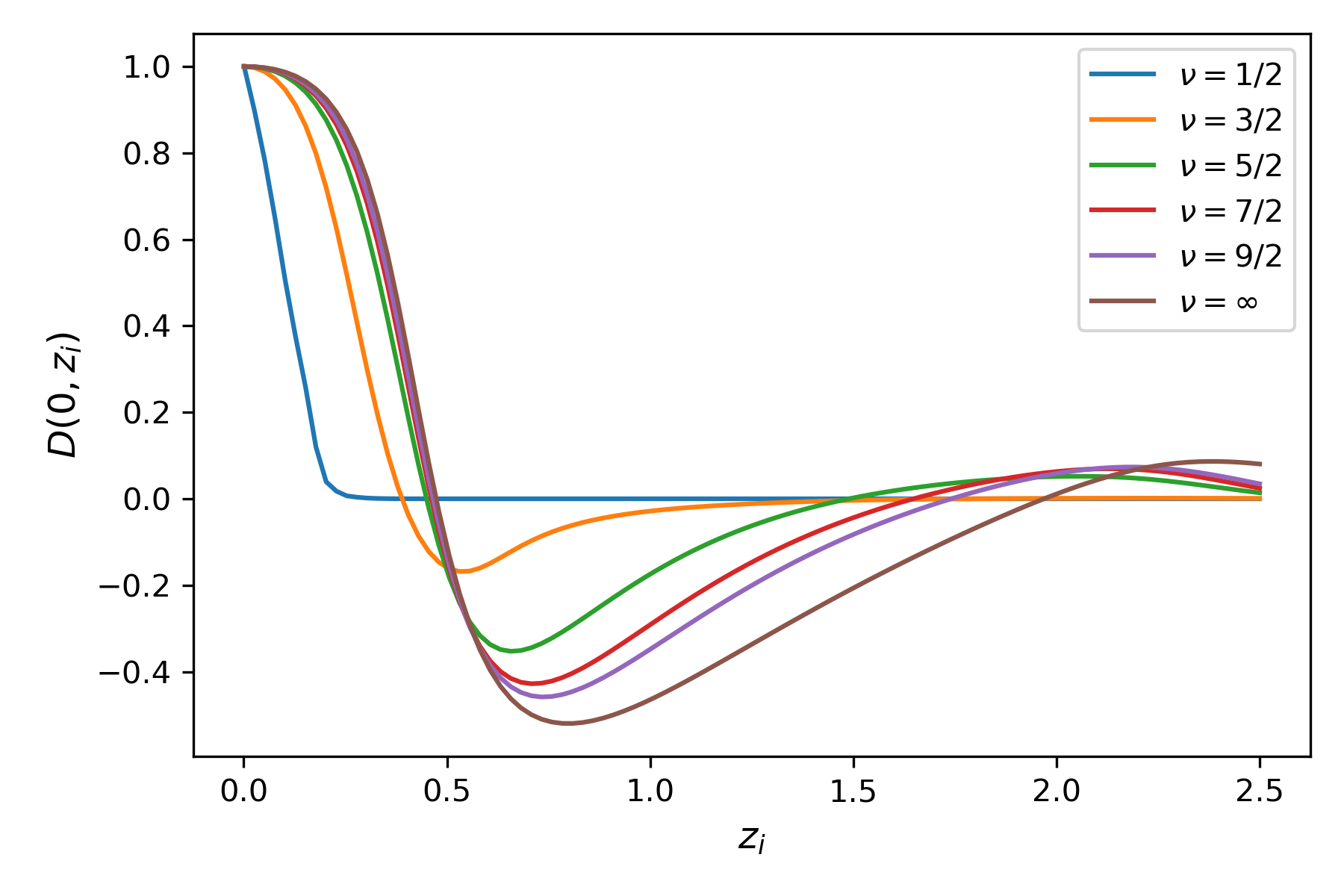} \\
\caption{Correlations between $H(z=0)$ and $H(z_i)$ across a host of Mat\'ern class kernels.}
\label{kernelD}
\end{figure}

Next, one can fix $z_i = 0$ and the first row of the correlation matrix gives us an indication of the correlations between $H(z =0)$ and $H(z_i)$. These are shown in FIG. \ref{kernelD}, where it is clear that the $\nu = \frac{1}{2}$ kernel has the weakest correlations with redshift, whereas the Gaussian kernel ($\nu = \infty$) exhibits the strongest. It can also be observed that beyond $\nu = \frac{3}{2}$, the differences in the correlations are not so pronounced. This explains not only the trend in the $H_0$ errors in Table \ref{H0kernel}, but also why the difference in the $H_0$ errors beyond $\nu = \frac{3}{2}$ is not so great. It is an undeniable fact that stronger correlations lead to smaller errors.

One can then 
extend this analysis to parametric models, so that a direct comparison can be made. In fitting parametric models, one typically ends up with an MCMC chain for the parameters, which within the two-parameter family of dynamical dark energy (DDE) models we study, amounts to a maximum of four parameters $(H_0, \Omega_{m0}, w_0, w_a)$. Concretely, we make use of a redshift model \cite{Cooray:1999da, Astier:2000as}, the CPL model \cite{Chevallier:2000qy, Linder:2002et}, as well as models due to Efstathiou \cite{Efstathiou:1999tm}, Jassal-Bagla-Padmanabhan (JBP) \cite{Jassal:2005qc} and Barboza-Alcaniz (BA) \cite{Barboza:2008rh}.  As observed recently in \cite{Colgain:2021pmf}, focusing on any particular DDE parametrisation risks biasing the search for DDE, so here we analyse a broad class of models. We will see that our conclusions are robust to the parametrisation. 

From the MCMC chain, one can infer $H(z_i)$ at the same redshifts as the GP. From there one can make a direct comparison by plotting $H(z)$ and the confidence intervals. However, since the mean values may be displaced, it may be difficult to quantify the difference through plots. Nevertheless, one can boil any distinction down to numbers. Viewing $H(z_i)$ as parameters in their own right, one can infer the corresponding covariance matrix and strip away the errors to leave the correlation matrix. In FIG. \ref{modelD} we plot the correlations across the parametric models.

\begin{figure}[htb]
\centering
\includegraphics[angle=0,width=80mm]{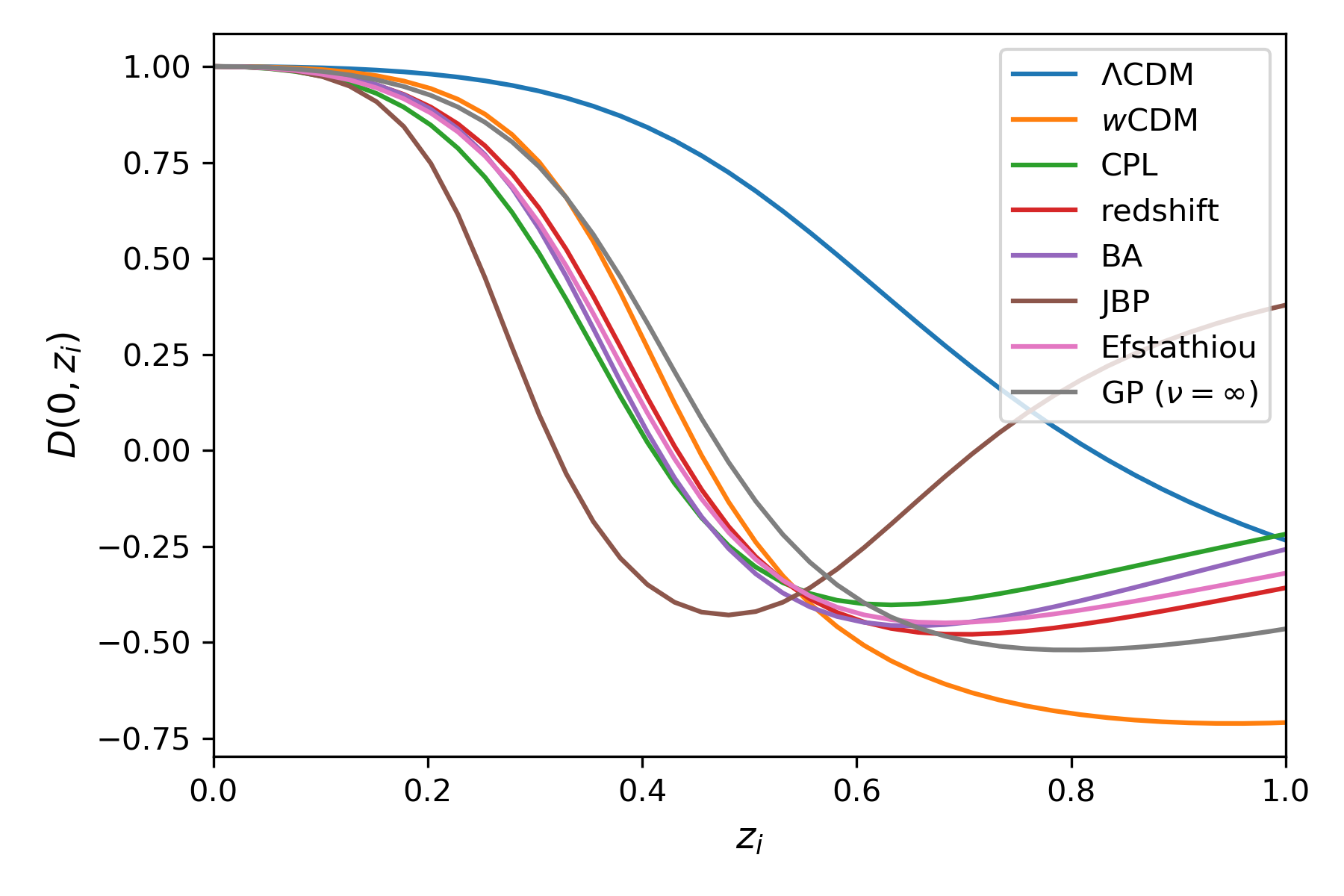} \\
\caption{Correlations between $H(z=0)$ and $H(z_i)$ across a host of DDE models. The Gaussian kernel is added for comparison.}
\label{modelD}
\end{figure}

In line with expectations, the flat $\Lambda$CDM model exhibits the strongest correlations, next $w$CDM, while any of the DDE models, including the CPL model, are less strongly correlated. This is more or less the content of Table \ref{param}, though, over a large number of mocks; here we only use the real data in FIG. \ref{data}. Ultimately, since the data is the same, these correlations are simply an artifact of the number of parameters in the model. However, for DDE models, the correlations are also in line with observations made in \cite{Colgain:2021pmf}, which is yet another consistency check. There it was noted that the errors on $w_a$, which propagate to all parameter errors, including $H_0$, are larger for the JBP and CPL models, while the Efstathiou, BA and redshift models lead to smaller errors. Once again, this is evident from FIG. \ref{modelD}. As explained the figures are only for a single realisation of the data, while Tables \ref{param} and \ref{nonparam} represent repeated mock realisations, so the conclusion that the errors on $H_0$ differ should be beyond doubt. 


\section{Conclusions}\label{sec:conclusion}
``Model independence" has casually slipped into the cosmology lexicon. Here, our interest in the claims were piqued by smallish errors in $H_0$ \cite{Busti:2014dua, Yu:2017iju, Gomez-Valent:2018hwc, Haridasu:2018gqm, Bonilla:2020wbn, Renzi:2020fnx}.\footnote{See \cite{Zhang:2021tmg} for a model independent $H_0$ determination with errors that raise few suspicions.} We started with some observations on Taylor expansion and the regime where it may be regarded as model independent in a \textit{bona fide} sense. As explained, these observations are rooted in $18^{\textrm{th}}$ and $19^{\textrm{th}}$ century math theorems, but this is routinely overlooked with dire consequences. Importantly, beyond the radius of convergence $|z|=1$, higher order terms in any expansion no longer converge. 

We also explained the difference between the two schools in the GP method \cite{Shafieloo:2012ht, Seikel:2012uu}. Clearly, an assumption on the mean function, as advocated in \cite{Shafieloo:2012ht}, can lead to very different expressions at leading order. In particular, a competitive guess on the mean, not only represents extra modeling, but risks triviliasing GP. Moreover, one can add a ``nugget" or noise contribution to $K+C$ \cite{Holsclaw:2010nb, Shafieloo:2012ht}, but this again is extra modeling. While these observations may not settle the debate, we believe they constitute progress. 

We analysed the correlations relevant for the inferred $H_0$ values with $\mu(z) = 0$ \cite{Seikel:2012uu}. Consistent with observations on the errors, we found that GP leads to stronger correlations and thus smaller errors than any well known parametric DDE model. If this is confirmed by the community, one must conclude that GP analysis is tantamount to fitting the $w$CDM model to the same data. While this may seem counter-intuitive, it is worth noting that GP has a myriad of applications, but in cosmology, $H(z)$ is to first approximation a rather dull monotonically increasing function of redshift. Therefore, it is possible that the optimum kernels for cosmology are yet to be identified. 

Finally, let us emphasise again that we have only analysed $H_0$. Our motivation was that competing discrepant ``model independent" $H_0$ determinations immediately lead to the conclusion that Hubble tension has no resolution in an FLRW cosmology. It is imperative to extend our results to other cosmological parameters, e. g. \cite{Benisty:2020kdt, GP-Omegak}, to ascertain the level of model dependence in the GP approach. It is worth recalling again our comments on Taylor expansions, namely that a Taylor expansion beyond the strict vicinity of $z \sim 0$ corresponds to a \textit{class of models}. In essence, GP should be the same. The responsibility is on the GP community to properly define the class of models in a transparent manner. 

\section*{Acknowledgements}
We thank Stephen Appleby and Tao Yang for discussions. We are also grateful to David Benisty, Chris Clarkson, Adri\`a G\'omez-Valant, Sandeep Haridasu, Eric Linder,  Rafael Nunes and Fabrizio Renzi for sharing their expertise on GP through helpful comments on the draft. E\'OC is funded by the National Research Foundation of Korea (NRF-2020R1A2C1102899). MMShJ would like to acknowledge SarAmadan grant No. ISEF/M/99131.

\appendix 

\section{Comments on Taylor Expansions}
\label{moreTE}
In order to make use of Taylor expansion as a ``model independent" approach, the practical problem is that $H(z)$ is determined by data and \textit{ab initio} it is unknown. To overcome this, one can naively expand $H(z)$, but one needs to decide on (i) the order $n$ and (ii) how far one can venture from $z=0$ so that the remainder function remains small. Some may regard this as an ``ad hoc" choice, however, it is relatively easy to address these two points once one demands that the standard model, flat $\Lambda$CDM, is covered by the expansion, i. e.  the expansion explores models close to $\Lambda$CDM.  
Since $H_0$ is an overall factor in $H(z)$,  it is enough to consider the normalised Hubble parameter $E(z) := H(z)/H_0$, which for flat $\Lambda$CDM is, 
\be
\begin{split}
E_{\textrm{exact}}(z) &:= \sqrt{1- \Omega_{m0} + \Omega_{m 0} (1+z)^3} \\ &\approx 1 + \frac{3\Omega_{m 0}}{2} z + \frac{3\Omega_{m0}}{8} ( 4 - 3 \Omega_{m 0} ) z^2 + \dots = E_{\textrm{approx}}(z),  \nonumber
\end{split}
\ee
where dots denote omitted terms. Note, we have not yet specified the order of the expansion. $E_{\textrm{exact}} (z)$ in the first line is the exact expression for flat $\Lambda$CDM whereas,  the expression on the second line, dubbed as $E_{\textrm{approx}}(z)$,  is a low $z$ approximation. Observe that for $z\gg 1$, $E_{\textrm{exact}}(z)\sim z^{3/2}$, which is a non-integer power of $z$, and that the coefficients of the $z, z^2, z^3$ (and in fact all the higher powers in the expansion) in $E_{\textrm{approx}}(z)$ are not independent and are specified in terms of $\Omega_{m 0}$. 

Let us begin by specifying the order $n$ and restrict our analysis to  $n \leq 10$, essentially to mimic analysis on higher order polynomials in Figure 6 of  \cite{Gomez-Valent:2018hwc}. Next, let us define the fractional difference between the exact expression for $E(z)$ and its approximation at a given $z$: 
\be
\label{dE}
\Delta E (z) = \frac{[E_{\textrm{exact}} (z) - E_{\textrm{approx}}(z) ]}{E_{\textrm{exact}}(z)}. 
\ee
We also introduce the discrete sum, 
\be
\label{S}
S(z_1, \dots z_{N}) = \sum_{i=1}^{N} [E_{\textrm{exact}} (z_i) -E_{\textrm{approx}}(z_i) ]^2, 
\ee
where we take $z_1 = 0$ and $\Delta z_i = z_i-z_{i-1} = 0.001$ with $z_{N}$ corresponding to the maximum value of the redshift range, which we rename $z_{\textrm{max}}$. Note that (\ref{dE}) is defined at a given redshift $z$, but (\ref{S}) is a summed quantity over a range of redshifts. For this reason, (\ref{S}) is a better measure of how close $E_{\textrm{approx}}(z)$ is to $E_{\textrm{exact}} (z)$. One can use the two measures interchangeably, since  they usually lead to the same conclusions, but if one wants to be more precise, the most accurate polynomial at a given order $n$ is the one that minimises (\ref{S}). It is worth emphasising again that the analysis here is purely analytical with no input from data.

The initial results of this exercise are shown in Table \ref{table2} ($\Omega_{m0} = 0.3$), where we have employed (\ref{S}) as a measure of precision. Simply put, polynomials with smaller $S$ numbers are more accurate, so this provides an easy way to rank the polynomials. Recall that the radius of convergence is at most $|z|=1$ \cite{Cattoen:2007sk}. For this reason, below $z_{\textrm{max}}=1$ one should expect that including higher order terms in the expansion will increase the agreement with the exact result. This is clearly the case when $z_{\textrm{max}} = 0.5$. Nevertheless, at $z_{\textrm{max}}=1$ and beyond,  adding higher order terms does not improve the approximation. This is evident from the $z_{\textrm{max}} =2.5$  entry, where the polynomials of lower order that perform the best. 

\begin{table}[htb]
\centering
\begin{tabular}{c|c|c}
\rule{0pt}{3ex} $z_{\textrm{max}}$ & Most precise $n$ & maximum $|\Delta E(z)|$ \\
\hline
\hline 
\rule{0pt}{3ex} $2.5$ & $3, 2, 5, \dots $ & $ 13.9 \%, 16.9 \%, 28 \%, \dots $ \\
\rule{0pt}{3ex} $2$ & $5, 3, 8, \dots $ & $ 8.7 \%, 9.2 \%, 16 \%, \dots $ \\
\rule{0pt}{3ex} $1.5$ & $10, 5, 8, \dots $ & $ 1.5 \%, 1.7 \%, 2.4 \%, \dots $ \\
\rule{0pt}{3ex} $1$ & $10, 5, 8, \dots $ & $ 0.008 \%, 0.1 \%, 0.13 \%, \dots $ \\
\rule{0pt}{3ex} $0.5$ & $10, 9, 8, \dots $ & $ 0.000016  \%, 0.00026 \%, 0.00056 \%, \dots $ \\
\end{tabular}
\caption{The precision of different order ($n \leq 10$) polynomials in recovering flat $\Lambda$CDM with $\Omega_{m0} = 0.3$ up to redshift $z_{\textrm{max}}$. We highlight only the 3 best performing polynomials.}
\label{table2}
\end{table}

This may be a little counter-intuitive, but the notion that \textit{higher order terms improve precision} is only true within the radius of convergence. 
These problems with convergence can be solved by expanding in the $y$-parameter, $y=z/(1+z)$, as advocated in \cite{Cattoen:2007sk}. The $y$-parameter, however, performs worse than $z$ below $z_{\textrm{max}}=1$, and one typically requires a large number of expansion parameters. See Figure 9 of \cite{Yang:2019vgk}. 

It is also worth noting from Table \ref{table2} that overall the $n=10$ polynomial only performs marginally better than $n=5$, and given that the latter has fewer parameters, this singles it out as a better choice. In summary, one can safely jettison the higher order terms $n > 5$ and it is sufficient to discuss Taylor expansion in $z$ up to fifth order. If one restricts attention to the requirement that $E_{\textrm{approx}}(z)$ recovers $E_{\textrm{exact}}(z)$ to within $1 \%$ error in a range of matter densities $0.25 \leq \Omega_{m0} \leq 0.35$ with the flat $\Lambda$CDM model, the $n=2$, $n=3$, $n=4$ and $n=5$ Taylor expansions are valid to $z_{\textrm{max}} \approx 0.66$, $z_{\textrm{max}} \approx 0.8$, $z_{\textrm{max}} \approx 0.86$ and $z_{\textrm{max}} \approx 1.14$, respectively. One can push the redshift cut-off higher, but this leads to poorer approximation. Note that the $n=4$ expansion is only marginally better than $n=3$, so this choice may be optimal below $z \sim 0.8$. 

\section{Mat\'ern Covariance Functions}
\label{App:Matern} 
For $\nu = p + \frac{1}{2}, p \in \mathbb{N}^+$, the Mat\'ern covariance matrix can be written as a product of an exponential and a polynomial of order $p$. Here we record some simplified expressions: 
\begin{widetext}
\bea
\label{K}
K_{1/2} (z, \tilde{z}) &=& \sigma_f^2 \exp \left( - \frac{ |z - \tilde{z}|}{\ell_f} \right),  \nn
K_{3/2} (z, \tilde{z}) &=& \sigma_f^2 \exp \left( - \frac{ \sqrt{3} |z - \tilde{z}|}{\ell_f} \right) \left( 1 + \frac{\sqrt{3} |z - \tilde{z}|}{\ell_f} \right),  \nn
K_{{5}/{2}} (z, \tilde{z}) &=& \sigma_f^2 \exp \left( - \frac{\sqrt{5} |z - \tilde{z}|}{\ell_f} \right) \left(1 + \frac{\sqrt{5} |z- \tilde{z}|}{\ell_f} + \frac{5 (z-\tilde{z})^2}{3 \ell_f^2} \right), \nn
K_{{7}/{2}} (z, \tilde{z}) &=& \sigma_f^2 \exp \left( - \frac{\sqrt{7} |z - \tilde{z}|}{\ell_f} \right)  \biggl(1 + \frac{\sqrt{7} |z- \tilde{z}|}{\ell_f} + \frac{14 (z-\tilde{z})^2}{5 \ell_f^2}  + \frac{7 \sqrt{7} |z - \tilde{z}|^3}{15 \ell_f^3} \biggr), \nn
K_{{9}/{2}} (z, \tilde{z}) &=& \sigma_f^2 \exp \left( - \frac{3 |z - \tilde{z}|}{\ell_f} \right) \left(1 + \frac{3 |z- \tilde{z}|}{\ell_f} + \frac{27 (z-\tilde{z})^2}{7 \ell_f^2} + \frac{18  |z - \tilde{z}|^3}{7 \ell_f^3} + \frac{27 (z-\tilde{z})^4}{35 \ell_f^4} \right).  
\eea
\end{widetext}

\end{document}